# Quantum Chinese Magic box

Radel Ben-Av

Azrieli Academic College Of Engineering Jerusalem,

26 Shreibom st, Israel 9103501

**Abstract** : This work introduces a new concept of "Chinese Magic Box". The general idea is to have a box such that the sender can store information in multiple drawers. The receiver is free to open any drawer. However, once the receiver opens the drawer – he can retrieve the information from that drawer only, that is, the information that was stored in the other drawers is lost. This property is achieved by storing the information using a set of non-orthogonal quantum states. The different "drawers" are realized by different orthogonal set of basis for the measurement. Once the measurement is performed, the information in this basis is retrieved. At the same time, due to wave function collapse the information in the other basis is lost and cannot be retrieved. I show how to construct a set of states for a **<u>single qubit</u>** to implement a "Box" with two or three "drawers". Some applications are discussed. Among them is a new non-symmetric Quantum Key Distribution with only single direction classical communication.

## 1  Background

The seminal paper on the use of quantum states for secure communication by S. Wiesner sparked a flow of research in this field. In that paper [1] Wiesner introduced the concept of storing information using conjugate coding – i.e. the information is stored in one out of two complementing set of states. One can retrieve the information only by reading from the correct channel. If the incorrect channel is used then the information can not be read and is lost. It was soon realized that this idea could be used for secure key distribution by the BB84 [2] protocol.

The main properties that are being used are the no-cloning [5] theorem together with the fact that the qubit is characterized by two complex numbers rather than a single number that specify the probability to find the bit in a state of "0" or "1". Thus, the qubit can be prepared in many different basis. The classical "bit" can be coded by a qubit either in the computational basis {|0>,|1>} or in the {|+>, |->} basis at the will of the preparer. The receiver of the qubit, can uncover the bit only if she knows (or can correctly guess) at which basis the bit was encoded. If the reader is guessing incorrectly – the result is zero information, that is – it has zero correlation with the stored information. In such a case, in addition to the zero correlation with the stored information - the stored information is destroyed due to wave function collapse. The no-cloning theorem ensures that the receiver cannot make two choices for the same qubit, unlike classical mechanics – where the reader is able to poke the systems as many time as she wishes with negligible effect on the system.

This phenomenon was then used by Bennet and Brassard (the BB84 protocol) to yield a Quantum Cryptographic protocol [2]. In that protocol - Alice (the sender) is generating a random code, part of which will be used as a cryptographic code. She then encodes the message randomly either in the computation basis ($|0>,|1>$) or the $\{|+>, |->\}$, basis. Bob who is on the receiving end of the communication will now also choose randomly at which basis to read. Both Bob and Alice publically share their choices of basis. Once Bob and Alice know their respective choice of basis they can establish a common sequence of random bits. A crucial aspect of the protocol is that one can detect whether there is an eve-dropper - Eve - in the loop.

In addition to this protocol, later Ekert devised a protocol based on EPR pairs [3,4]. Instead of sending qubits from Alice to Bob, an entangled pair is created such that the qubit that Alice receives is correlated with that of Bob. Like before, both Alice and Bob make their measurement and if they agree on the basis, they can be sure that the sequence of bits that they both measured is common and can be used as a secret code. The no-cloning theorem protects also this code from being copied without both Alice's and Bob's awareness. There are also many practical implementation of the Quantum Key Distribution [6-8].

In next section, I will first describe how a sender can store two different bits in a **single qubit**. The caveat being that only one of the bits can be retrieved (with some error probability), once the bit is retrieved – the other bit cannot be retrieved. Which channel is retrieved and which is destroyed is given to the choice of the reader.

I will than show how similar result can be achieved with three bits and a **single qubit**.

In section 3, I will provide possible applications and in particular a one-directional classical communication for quantum key distribution.

Section 4 will summarize the work and provide issues for further research.

## 2   Multi—Compartment Qubit

In this section I will show how to build magic boxes with two or three "drawers" using single qubit.

2.1   Two Drawers Box

Let us define the states $\Psi_{\alpha,\beta}$

$$\begin{aligned}\Psi_{0,0} &= a|0> +b|1> \\ \Psi_{0,1} &= a|0> -b|1> \\ \Psi_{1,0} &= b|0> +a|1> \\ \Psi_{1,1} &= b|0> -a|1>\end{aligned} \quad (1)$$

Let set $a = \cos\left(\frac{\pi}{8}\right) \quad b = \sin(\frac{\pi}{8})$.

Alice can store the two classical bits $(\alpha, \beta)$ in the same qubit by setting the state of the qubit to the appropriate $\Psi_{\alpha,\beta}$ state.

Alice will send the qubit to Bob.

Now if Bob wishes to retrieve the $\alpha$ bit – he will measure the qubit in the computational basis - {|0>, |1>}. It can be easily seen that the probability to get the correct result is :.

$$|a|^2 = \cos^2\frac{\pi}{8} = \frac{1}{2} + \frac{\sqrt{2}}{4} = 0.85355339059 \quad (2)$$

If Bob wishes to retrieve the $\beta$ bit he will measure the qubit in the {|+>, |->} basis.

Where

$$\begin{aligned}|+> &= \frac{1}{\sqrt{2}} \{|0> + |1>\} \\ |-> &= \frac{1}{\sqrt{2}} \{|0> - |1>\}\end{aligned} \quad (3)$$

If the result is |+> then Bob registers the value "0", and if the result is |-> than Bob registers the value "1". One can see that the probability to get the correct result is again

$$|a|^2 = \cos^2 \frac{\pi}{8} = 0.85355339059 \qquad (4)$$

After Bob's retrieving the α bit - the β bit is lost, since after measuring in the computational basis - the qubit will be in an eigenstate of the computational basis with equal probability for results in the {|+>, |->} basis, and vice versa.

The information content of the qubit is be calculated by noticing that each "drawer" can be considered as a simple binary channel with random noise. Using Shannon's information theory [9] one gets that the information content of the α bit is:

$$I = 1 + p \log_2(p) + (1 - p) \log_2(1 - p) = 0.39912396329 \qquad (5)$$

The same calculation applies for the β bit. Hence, the total amount of information carried by each qubit is 2*I ≅ 0.8. This value is clearly smaller than 1 but not significantly. At the same time, the information that can actually be retrieved from each qubit is only ~ 0.4 bits. The choice of which "drawer" to retrieve is in the hands of the reader. Once the qubit has been read the information in the other drawer is lost. It is also clear that this process does not violate the Holevo bound [10].

Note: Given the above derivation one can show that the choice of value for a is optimal in the sense of minimal number of qubits sent from Alice to Bob assuming that the same amount of information is carried in the computational and the {|+>,|->} basis.

## 2.2  Three Compartments Box

Let us define the states $\Psi_{\alpha,\beta,\gamma}$

$$\begin{aligned}
\Psi_{0,0,0} &= a\,|0> + e^{\frac{i\pi}{4}} b\,|1> \\
\Psi_{0,0,1} &= a\,|0> - i e^{\frac{i\pi}{4}} b\,|1> \\
\Psi_{0,1,0} &= a\,|0> + i e^{\frac{i\pi}{4}} b\,|1> \\
\Psi_{0,1,1} &= a\,|0> - e^{\frac{i\pi}{4}} b\,|1> \qquad (6)\\
\Psi_{1,0,0} &= b\,|0> + e^{\frac{i\pi}{4}} a\,|1> \\
\Psi_{1,0,1} &= b\,|0> - i e^{\frac{i\pi}{4}} a\,|1> \\
\Psi_{1,1,0} &= b\,|0> + i e^{\frac{i\pi}{4}} a\,|1>
\end{aligned}$$

$$\Psi_{1,1,1} = b\left|0\right> - e^{\frac{i\pi}{4}}a\left|1\right>$$

Let us now define the standard three basis of the qubit (corresponding to the three eigensvectors of the Pauli matrices):

$B_1$ = {|0>, |1>)    $B_2$ = {|+>, |->}   and $B_3$ = {|u>, |d>}   where

$$|+> = \frac{1}{\sqrt{2}}\{|0> + |1>\}$$
$$|-> = \frac{1}{\sqrt{2}}\{|0> - |1>\}$$
$$|u> = \frac{1}{\sqrt{2}}\{|0> + i|1>\} \quad\quad (7)$$
$$|d> = \frac{1}{\sqrt{2}}\{|0> - i|1>\}$$

Whenever Alice wishes to transfer a triplet of bits – $(\alpha, \beta, \gamma)$ to Bob, she will construct the state $\Psi_{\alpha,\beta,\gamma}$ and transfer it to Bob. Bob can choose which channel to retrieve. If he wishes to retrieve the $\alpha$ channel, he will measure in the basis $B_1$ with obvious interpretation. If he wishes to retrieve the $\beta$ channel, he will measure in the basis $B_2$, where |+> is interpreted as "0", and |-> as "1". If he wishes to retrieve the $\gamma$ channel, he will measure in the basis $B_3$, where |u> is interpreted as "0", and |d> as "1". In all cases Bob will retrieve the original bit with some error probability.

One can calculate the probability of the results in all three bases – $B_1$, $B_2$ and $B_3$. Table 1 summarizes the results of measuring the above-defined states in any of the above-mentioned bases.

|  | $B_1$ | $B_2$ | $B_3$ |
| --- | --- | --- | --- |
| $\Psi_{0,0,0}$ | Prob(|0>) = p | Prob(|+>) = q | Prob(|u>) = q |
| $\Psi_{0,0,1}$ | Prob(|0>) = p | Prob(|+>) = q | Prob(|d>) = q |
| $\Psi_{0,1,0}$ | Prob(|0>) = p | Prob(|->) = q | Prob(|u>) = q |
| $\Psi_{0,1,1}$ | Prob(|0>) = p | Prob(|->) = q | Prob(|d>) = q |
| $\Psi_{1,0,0}$ | Prob(|1>) = p | Prob(|+>) = q | Prob(|u>) = q |
| $\Psi_{1,0,1}$ | Prob(|1>) = p | Prob(|+>) = q | Prob(|d>) = q |
| $\Psi_{1,1,0}$ | Prob(|1>) = p | Prob(|->) = q | Prob(|u>) = q |

| $\Psi_{1,1,1}$ | Prob($|1\rangle$) = p | Prob($|\text{->}\rangle$) = q | Prob($|d\rangle$) = q |
|---|---|---|---|

Table 1 – Summary of measurement results

Where

$$p = a^2$$
$$q = \frac{1}{2} + \frac{\sqrt{2}}{2}\sqrt{p(1-p)} \qquad (8)$$

If p>0.5 than the probability of getting the correct answer is also $P_{correct}$> 0.5. Hence one can consider this scheme for each channel ("drawer") also as a classical noisy channel.

We can choose p such that q=p - achieving complete symmetry between all three channels (this is also probably also the optimal choice). This condition is fulfilled when

$$p = \frac{1}{2} + \frac{\sqrt{12}}{12} \qquad (9)$$

In this case, the amount of information per channel can again be calculated using Shannon's information formula:

$$I = 1 + p \log_2(p) + (1-p) \log_2(1-p) = 0.2559924488 \qquad (10)$$

That is each channel contains approximately ¼ bit. The overall amount of information stored in all the tree channels is 3*I ≅ 0.76.

Sending a sequence of N bits in each of the "drawers" can be achieved using N/0.39912396329 qubits in the two "drawers" case and N/0.2559924488 qubits in the three "drawers" case.

Hence I have shown how to store two or three (fractional) bits in the same qubit. The receiver can retrieve either of them. Once a bit has been retrieved, the bits in the other "drawers" are not accessible.

# 3   Applications

In this part, I will describe two possible application for the above mechanism. The first application is a direct use of the "magic-box". The second application is a new Quantum Key Distribution Protocol (QKD). The advantage of the new QKD protocol is that classical communication is required only in one direction.

## 3.1   Application to Merchandize

Suppose that a distributor of a digital data wishes to distribute a set of two digital contents (like multi-video series or two books etc) to a customer. In addition - suppose that he wants to grant the customer a license to access only one of the two possible contents. One possible solution is to decrypt both contents with different keys. The receiver of the contents will be required to access the distributor in order to receive a key for any of the possible contents. Using the above mentioned CMB (Chinese Magic Box) – the sender may encode the content that she wishes on one of the channels (i.e. a sequence of Quantum CMB's) and the second content on the other channel. An N-bits content requires N/0.3991qubits, due to the nosiness of the channel. The advantage is that the distributor is assured that the receiver will have access only to one of the contents. Once the receiver retrieved the content that he wishes – the other content is not available.

Using the three compartment CMB one can achieve the same goal for three possible contents in a similar manner. Notice that the total overhead in qubits is only 25%.

### 3.2  Application to Quantum Key Distribution

In this section, I will describe a QKD protocol using CMB.

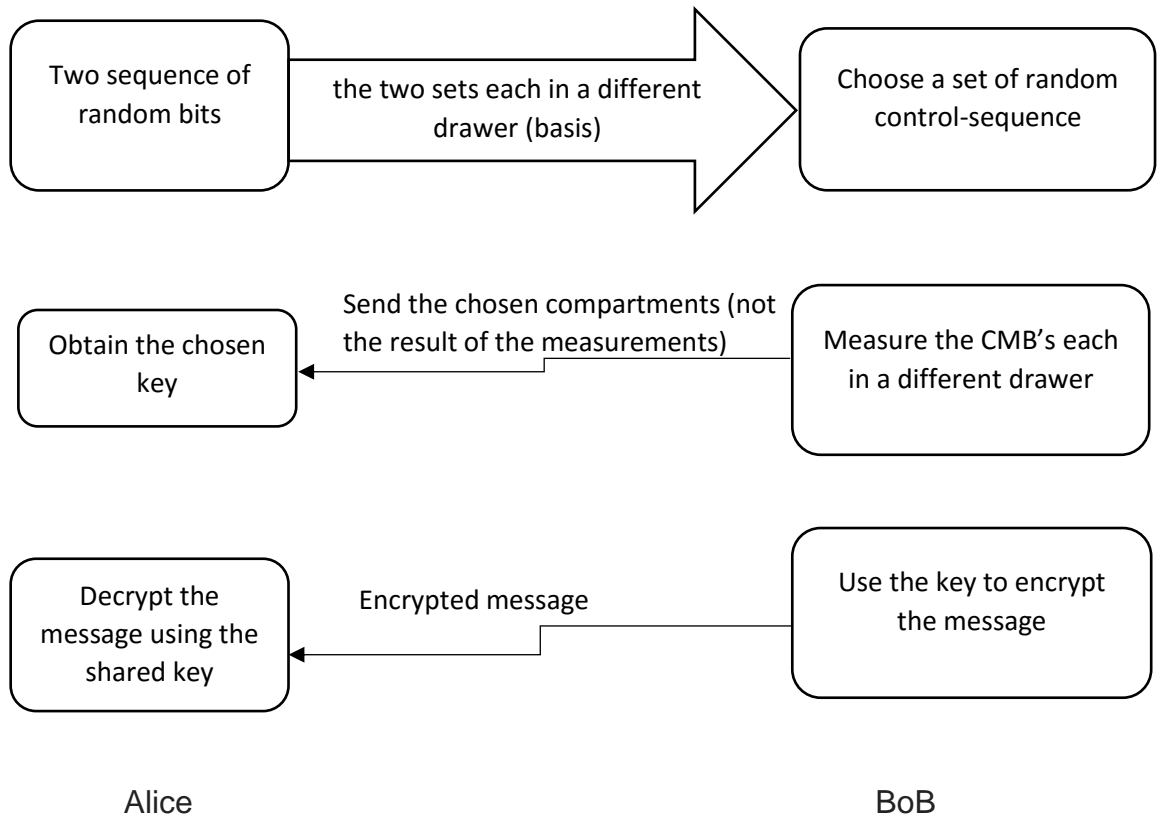

Figure 1 - Asymmetric QKD Protocol

Suppose that Alice sends a set of two-drawer Magic Boxes to Bob. Each drawer in every CMB is filled with a random bit. At this stage we will ignore the extra CMB's needed due to the noisiness of the CMB's. I will address this point in appendix A. Alice will register her CMB contents. Upon receiving the CMB's Bob will choose another random sequence of 1's and 0's – the control sequence. If the bit in the control sequence is 0(1) he will read from drawer 1(2), respectively. Bob will then have a set of random bits read from the CMB's –the key sequence. He will then use this key-sequence of bits as a key to encrypt his message. In order to share his key with Alice, Bob will send his random control sequence openly to Alice. He can than send his encrypted message to Alice using open communication. Since Alice knows the content of her CMB's she can reconstruct the encryption-key based on the open control-sequence and thus she can decrypt the message from Bob. Since the control-sequence has no correlation to the key-sequence, nobody else can deduce the key-sequence other than Ellis. This process is shown in figure 1.

Notice that the CMB's sending can be done with significant time ahead of the classical communication. In this protocol there is only classical communication from Bob to Alice. Alice need not send or publish any classical communication. This is an advantage with respect to the BB84 protocol, where there is a need for classical communication from Alice to Bob as well as from Bob to Alice in order to negotiate on the agreed code. On the other hand, in the BB84 protocol, there is a need for 2 qubits on average for every single bit of code. In the CMB protocol there is a need for much more qubits per single code bit (see appendix A).

## 4    Summary

In this paper, I have shown how to construct a two or three drawers Quantum Chinese Magic Box. The property of the CMB is that one can open/read only one drawer. Once this drawer has been accessed, the content of the other drawers is not accessible. The CMB is constructed of a single qubit and it contains two or three "drawers" each with a fractional bit. This construction is based upon the wave function collapse and the no-cloning theorem.

I have described two possible applications for the CMB. The first is related to content distribution where the distributor can control the amount of received data without a need for post distribution communication. The second application is a new QKD distribution protocol where there is a need only for one directional classical communication. This can be very helpful when one side of the communication cannot send information but only receive it, for example when its location is not known or when it does not have enough power for sending information. It may also require less time for the classical communication since only one direction communication is required.

The following questions remain for future work – is it the most efficient CMB? Is there any advantage in using qudits as a basis for CMB? How does the CMB function when there is a quantum noise?

**Acknowledgements**

Thanks for many helpful discussions to Wei Shijie Wei, Quan Quan and Ma Teng from Prof. G. Long's group and to Prof. R. Ben-Eliyahu Zohary.

# Appendix A

In this appendix, I will describe a protocol for generating a shared sequence of K random bits between Alice and Bob, using CMB's and taking into account the noisiness of the reading. The protocol will also involve steps that verify that the key was not compromised by an eve-dropper.

**Alice Preparation** - Alice generates two random sequence of T bits (T ≥ K). One can think of it as a sequence of T <u>pairs</u> of random bits. Alice keeps this sequence of T-pairs. In addition, Alice will prepare T blocks of R qubits. Each block will correspond to one pair. All the qubits in the same block will be identical and contain the first bit of the pair in the first "drawer" and the second bit in the second "drawer". Alice will send the T*R qubits to Bob.

**Bob Preparation and Encryption** – Bob generates a sequence of T random control bits. Each qubit in the same block is measured using the same basis. Different blocks can be measured using different basis. The value of the bit represented by the block is determined using the majority rule.

Bob will randomly choose K blocks out of the T blocks. Bob will use the results of his measurements of these K blocks as his random-key. The results of all the T blocks will be used for eve-dropping detection (see later).

If no eve-dropping was detected - Bob will share his choice of K-blocks together with his control bits for these blocks with Alice in an open channel.

Bob will use his sequence of K bits as a key to encrypt his message and will send his encrypted message to Alice.

**Alice Decryption** – Having receive the K-blocks id's and the K control sequence from Bob, Alice can reconstruct the sequence of K bits that Bob uses as his encryption key. If there was no eve-dropping than only Alice can know what is the K-bits encryption key. She can decrypt the message from Bob using this key.

**CMB Noise and Choosing R** - Clearly the larger the value of R is, the higher is the probability of Bob to retrieve the value that Alice used for constructing the CMB's. For any $\epsilon > 0$ one can find R such that the probability of Bob to retrieve correctly all the K-bits is higher than $1 - \epsilon$

**Eve Dropping and Eve Dropping Detection** – Assume that there is an Eve with access to the T*R qubits between the time that Alice generated them and the time that Bob is using them. Since she does not know in advance the control sequence, she cannot measure them in the basis that Bob will measure them. However, due to the noisiness of the channel each block contains R>2 qubits. Eve may use this fact in order to measure two qubits in each block. The first qubit will be measured in the computational basis and the second in the {|+>,|->} basis. Since Bob measures each block in only one of the basis – Eve may have approximately $p = \cos^2 \frac{\pi}{8} = 0.85355$ correlation with Bob's result. In order to detect this occurrence of this scenario, Bob can check the distribution of the results in his blocks. If there was no eve-dropping the distribution is governed by the original states in the following way.

The measurement results of Bob will be denoted by $M_{ij}$ where $1 \leq i \leq T$ is the block number and $1 \leq j \leq R$ is the index of the qubit in the block. Bob's decision about each block will be denoted by $B_i = majority_j(M_{ij})$. Define

$$E = \frac{\Sigma_{i,j}(M_{ij} - B_i)^2}{R \cdot T}$$

If there was no eve-dropping than $<E> = p = 0.85355$. However if Eve has measured two qubit per block then $<E> = p_e = 0.8535 - \frac{0.5}{R}$. For any given R and for any $\epsilon > 0$ one can find $T_0$ such that for any $T \geq T_0$ the probability $P[E < (p + p_e)/2]$ is higher than $1 - \epsilon$ in case when Eve chooses to measure two qubits in every block, each in a different basis. Using $T = T_0$ Bob can measure E and he can validate whether there was eve-dropping or not by measuring E. If $E < (p + p_e)/2$ he will deduce that there was an eve-dropping.